\documentclass{ifacconf}
\usepackage{amsmath} 
\usepackage{amssymb}  
\usepackage{graphicx}      
\usepackage{natbib}        
\usepackage{amsmath}
\usepackage{color}
\usepackage{float}
\usepackage{bm}

\newcommand{\matr}[1]{\bm{#1}}

\renewcommand{\S}{\matr{S}} 
\newtheorem{lemma}{Lemma}

\newtheorem{theorem}{Theorem}
\newtheorem{remark}{Remark}
\newtheorem{definition}{Definition}

\begin{document}
\begin{frontmatter}

\title{Implicit Nonholonomic Mechanics with Collisions\thanksref{footnoteinfo}} 

\thanks[footnoteinfo]{ARA has been partially supported by Ministerio de Ciencia e Innovación (Spain) under grant PID2021-126124NB-I00. LC acknowledges financial support from Grant PID2022-137909NB-C21 funded by MCIN/AEI/ 10.13039/501100011033.}

\author[First]{\'Alvaro Rodr\'iguez Abella} 
\author[Second]{Leonardo Colombo}

\address[First]{Department of Mathematics and Computer Science, Saint Louis University (Madrid Campus), Avenida del Valle, 34, Madrid, 28003, Madrid, Spain. email: alvrod06@ucm.es}
\address[Second]{Centre for Automation and Robotics (CSIC-UPM), Ctra. M300 Campo Real, Km 0,200, Arganda
del Rey - 28500 Madrid, Spain. email:  leonardo.colombo@csic.es}

\begin{abstract}    
In this paper, variational techniques are used to analyze the dynamics of nonholonomic mechanical systems with impacts. Implicit nonholonomic smooth Lagrangian and Hamiltonian systems are extended to a nonsmooth context appropriate for collisions. In particular,  we provide a variational formulation for implicit nonholonomic mechanical systems with collisions, for those collisions that preserve energy and momentum at the impact. Lastly, the theoretical results are illustrated by examining the example of a rolling disk hitting a wall.
\end{abstract}

\begin{keyword}
Implicit Lagrangian systems, Implicit Hamiltonian systems, Nonholonomic systems, Lagrange--d'Alembert--Pontryagin principle, Collision dynamics.
\end{keyword}

\end{frontmatter}

\section{Introduction}
A nonholonomic system is a mechanical system subject to constraint functions which are, roughly speaking, functions on the velocities that are not derivable from position constraints.
They arise, for instance, in mechanical systems that have rolling or certain types
of sliding contact. There are multiple applications in the context of wheeled
motion, mobile robotics and robotic manipulation.

A geometrical formulation for mechanical systems with one-sided constraints was developed by Lacomba and Tulczyjew (see \cite{Lacomba1990}).
Ibort \textit{et al.} studied the geometric aspects of Lagrangian systems subject to impulsive and one-sided constraints  (\cite{ibort1998geometric}). This was extended to the Hamiltonian formalism by  \cite{cortes2006hamiltonian}. 

Mechanical systems subject to collisions are confined within a region of space with a boundary. Collision with the boundary for elastic impacts activates a constraint on the momentum and on the energy after and before the collision. The problem of collisions has been extensively treated in the
literature since the early days of mechanics (see \cite{brogliato1999nonsmooth} for a comprenshive review and references therein). More recently, much work has been done on the rigorous mathematical foundation of impact problems (see \cite{haddad2006impulsive} and \cite{westervelt2018feedback}). Nonholonomic systems subject to impacts or impulse effects have been previously studied in \cite{clark2019bouncing} and \cite{colombo2022geometric}.

In mechanics, implicit Lagrangian and Hamiltonian systems appear in controlled mechanical systems. An important class of implicit mechanical systems studied in \cite{yoshimura2006dirac} are those with nonholonomic constraints. The aim of this paper is to take one step further and consider implicit mechanical systems subject to nonholonomic constraints and elastic collisions, which occurs when the nonholonomic system impacts the boundary of the configuration space under some suitable conditions. The goal of this paper is to provide a variational formulation for nonholonomic implicit mechanical systems with collisions. In particular, those collisions that preserve energy and momentum at the impact.

The remainder of the paper is structured as follows. Section \ref{sec:main} introduces nonholonomic systems from an explicit point of view via the Lagrange--d'Alembert principle and from an implicit point of view via the Lagrange--d'Alembert--Pontryagin principle.  In Section \ref{sec3} we define the configuration space and the phase space with the objective of introducing the action functional in Section \ref{sec4}, where we derive variationally, via the Hamilton–-d’Alembert–-Pontryagin principle, the equations of motion for implicit nonholonomic Lagrangian systems subject to elastic collisions. We extend the framework to the Hamiltonian side in Section \ref{sec5}, where we derive nonholonomic implicit Hamiltonian systems subject to collisions from a variational perspective. Finally, we study the vertical rolling disk hitting a wall in Section \ref{sec6}.

\section{Nonholonomic systems}
\label{sec:main}
Let $Q$ be a differentiable manifold with $\hbox{dim}(Q)=n$. Throughout the text, $q^{i}$ will denote a particular choice of local coordinates on this manifold and $TQ$ denotes its tangent bundle, with $T_{q}Q$ denoting the tangent space at a specific point $q\in Q$. Usually $v_{q}$ denotes a vector at $T_{q}Q$ and, in addition, the coordinate chart $q^{i}$ induces a natural coordinate chart on $TQ$ denoted by $(q^{i},\dot{q}^{i})$ with $\hbox{dim}(TQ)=2n$. Let $T^{*}Q$ be its cotangent bundle, locally described by the positions and the momentum for the system, i.e., $(q,p)\in T^{*}Q$ with $\hbox{dim}(T^{*}Q)=2n$. The cotangent bundle at a point $q\in Q$ is denoted as $T_{q}^{*}Q$.


 A $k$-dimensional distribution $\Delta_Q$ on a manifold $Q,$ is a $k$-dimensional subspace $\Delta_Q(q)$ of $T_qQ$ for each $q\in Q$. $\Delta_Q$ is smooth if for each $q\in Q$ there exist a neighborhood $U$ of $q$ and $k$ $ C^{\infty}$ vector fields $X_1,\ldots,X_k$ on $U$ that span $\Delta_Q$ at each point of $U$, that is, $\Delta_Q(q)=\hbox{span}\{X_{1}(q),\ldots,X_k(q)\}$. 
 The \textit{rank of $\Delta_Q$ at $q\in Q$} is the dimension of the subspace $\Delta_Q(q)$, i.e. $\varrho:Q\to\mathbb{R},$ $\varrho(q)=\dim\Delta_Q(q).$ For any $q_0\in Q$ it is clear that $\varrho(q)\geq\varrho(q_0)$ in a neighborhood of $q_0.$ If $\varrho$ is a constant function, then $\Delta_Q$ is called a \textit{regular distribution}. In an analogous fashion as for distributions, it is possible to
define \textit{codistributions.} 

A \textit{smooth regular codistribution} $\widetilde{\Delta_Q}$ on
$Q$ is a subbundle of $T^{*}Q$ with $k$-dimensional fiber. Given the concept of codistribution, it is possible to define the \textit{annihilator} of a distribution. Let $\Delta_Q\subset TQ$ be a distribution, the annihilator of $\Delta_Q$ is the codistribution defined as
\[
\Delta_Q^\circ(q)=\{\alpha\in T_{q}^{*}Q\,\,|\,\,\alpha(v)=\langle\alpha,v\rangle=0,\,\,\forall v\in\Delta_Q(q)\}
\] for every $q\in Q$.

 Linear constraints on the
velocities are locally given by equations of the form
\[
\phi^{a}(q^i, \dot{q}^i)=\mu^a_i(q)\dot{q}^i=0, \quad 1\leq a\leq
m,
\]
depending, in general, on their configuration coordinates and  their
velocities. {}From an intrinsic point of view, the linear
constraints are defined by a regular distribution $\Delta_Q$ on
$Q$ of constant rank $n-m$ such that the annihilator of $\Delta_Q$ is locally given at each point of $Q$ by
\[
\Delta_Q^\circ(q) = \operatorname{span}\left\{ \mu^{a}(q)=\mu_i^{a}dq^i \; ; 1 \leq a
\leq m \right\},
\]
where the $1$-forms $\mu^{a}$ are independent at each point of $Q$.

We will restrict ourselves to the case of linear nonholonomic constraints.
In this case, the constraints are given by a nonintegrable
distribution $\Delta_Q$. In addition to these constraints, we
need to specify the dynamical evolution of the system, usually by
fixing a Lagrangian function $L\colon  TQ \to \mathbb{R}$. The
central concepts permitting the extension of mechanics from the
Newtonian point of view to the Lagrangian one are the notions of
virtual displacements and virtual work. These concepts were
formulated in the developments of mechanics and in their application to
statics. In nonholonomic dynamics,  the procedure is given by the
\textit{Lagrange--d'Alembert principle}.
 This principle allows us to determine the set of possible values of the constraint forces from the set $\Delta_Q$ of admissible kinematic states alone. The resulting equations of motion are
\begin{equation*}
\left[ \frac{d}{dt}\left( \frac{\partial L}{\partial \dot
q^i}\right) - \frac{\partial L}{\partial q^i} \right] \delta
q^i=0,
\end{equation*}
where $\delta q^i$ denotes the virtual displacements verifying
\begin{equation*}
\mu^a_i\delta q^i =0.
\end{equation*}
By using Lagrange multipliers, we
obtain
\begin{equation}\label{ldaeq}
\frac{d}{dt}\left( \frac{\partial L}{\partial \dot
q^i}\right)-\frac{\partial L}{\partial q^i}={\lambda}_a\mu^a_i  .
\end{equation}
The term on the right-hand side represents the  constraint force or
reaction force induced by the constraints and the functions $\lambda_a$ are the Lagrange multipliers which, after
being computed using the constraint equations, allow us to obtain a
set of second-order differential equations.

Alternatively to the use of Lagrange multipliers, the phase space may be enlarged to the Pontryagin bundle $TQ\oplus T^{*}Q$ and the Lagrange--d'Alembert--Pontryagin principle may be considered. This variational principle is given by $$\delta\int_{t_0}^{t_1}\left(L(q(t),v(t))+\langle p(t),\dot{q}(t)-v(t)\rangle\right)dt=0,$$
where $v(t)\in\Delta_Q(q(t))$ and the variations $(\delta q(t),\delta v(t), \delta p(t))$ are such that $\delta q(t)\in\Delta_Q(q(t))$ and vanishes at the endpoints. Then stationary condition for a curve $(q(t),v(t),p(t))$ yields the implicit Lagrange--d'Alembert equations on $TQ\oplus T^{*}Q$ (see \cite{yoshimura2006dirac}): $$p=\frac{\partial L}{\partial v},\quad\dot{q}=v\in\Delta_Q(q),\quad\dot{p}-\frac{\partial L}{\partial q}\in\Delta_Q^\circ(q).$$

\section{Configuration space and phase space}\label{sec3}

Let $Q$ be a smooth manifold with boundary, denoted by $\partial Q$, $L:TQ\to\mathbb R$ be a (possibly degenerate) Lagrangian, and $\Delta_Q\subset TQ$ be a (possibly nonholonomic) constraint distribution. 
According to Section \ref{sec:main}, the annihilator of $\Delta_Q$ is denoted by $\Delta_Q^\circ\subset T^*Q$.


 Given $[\tau_0,\tau_1]\subset\mathbb R$ and $\tilde\tau\in [\tau_0,\tau_1]$, the \emph{path space with a unique collision (at $\tau=\tilde\tau$)} is defined as $\Omega(Q,\tilde\tau)=\mathcal T\times\mathcal Q(\tilde\tau)$, where
\begin{equation*}
\mathcal T=\left\{\alpha_T\in C^\infty([\tau_0,\tau_1])\mid \alpha_T'(\tau)>0,~\tau\in [\tau_0,\tau_1]\right\}
\end{equation*}
and
\begin{align}\label{eq:mathcalQ}
&\mathcal Q(\tilde\tau)=\big\{\alpha_Q\in C^0([\tau_0,\tau_1],Q)\mid\alpha_Q(\tilde\tau)\in\partial Q,\\
&\alpha_Q\text{ is piecewise }C^2\text{ and has only one singularity at }\tilde\tau\big\}.\nonumber
\end{align}
We only consider one singularity at $\tau=\tilde\tau$ for brevity, but similar results hold for a finite amount of singularities, $\{\tilde\tau_i\mid 1\leq i\leq N\}\subset [\tau_0,\tau_1]$.

\begin{remark}Systems with collisions are a particular instance of hybrid systems. For systems with elastic impacts, the \emph{guard} is given by $S=\{v_q\in T_q Q\mid q\in\partial Q,~g(v_q,n_q)>0\}$, where $g$ is a Riemannian metric on $Q$ and $n$ is the outward-pointing, unit, normal vector field on the boundary. Similarly, the \emph{reset map} is given by $R(v_q)=v_q^{_\parallel}-v_q^{\perp}$, where $v_q^\perp=g(v_q,n_q)\,n_q$ and $v_q^{_\parallel}=v_q-v_q^\perp\in T_q\partial Q$. Recall that hybrid systems may experience Zeno behaviour if a trajectory undergoes infinitely many impacts in finite time. In order to avoid this situation, we ask the system to satisfy two conditions (cf. \cite[Remark 2.1]{GoCo2020}):

$(i)$ $S\cap\overline R(S)=\emptyset$, where $\overline R(S)$ is the closure of $R(S)\subset TQ$. This condition is clearly satisfied in our case. Indeed, for each $v_q\in S$ we have $v_q^\perp\neq 0$ and, thus, $\parallel R(v_q)-v_q\parallel_g=2\parallel v_q^\perp\parallel_g>0$, being $\parallel\cdot\parallel_g$ the norm induced by the metric $g$.

$(ii)$ The set of collision times is closed and discrete. This condition, which depends on the topology of the configuration manifold, prevents the existence of an accumulation point and will be assumed in the following.

    \end{remark}
Under these assumptions, our development is valid in a neighborhood of each collision.

\begin{lemma}{\cite[Corollary 2.3]{FeMaOrWe2003}}
$\Omega(Q,\tilde\tau)=\mathcal T\times\mathcal Q(\tilde\tau)$ is a smooth manifold.
\end{lemma}

\begin{remark}
Given $\alpha_T\in\mathcal T$, we denote $[t_0,t_1]=\alpha_T([\tau_0,\tau_1])$ and, in order to distinguish between $\tau$-derivatives and $t$-derivatives, we use different symbols; namely, $\alpha_T'=d\alpha_T/d\tau$ and $\dot\alpha_T^{-1}=d\alpha_T^{-1}/dt$, where $\alpha_T^{-1}:[t_0,t_1]\to[\tau_0,\tau_1]$ is the inverse of $\alpha_T$. Analogously, we denote $\tilde t=\alpha_T(\tilde\tau)$.
\end{remark}

The tangent space of $\mathcal Q(\tilde\tau)$ at $\alpha_Q\in\mathcal Q(\tilde\tau)$ is given by
\begin{align}\label{eq:TmathcalQ}
& T_{\alpha_Q}\mathcal Q(\tilde\tau)=\Big\{\nu_{\alpha_Q}\in C^0([\tau_0,\tau_1],TQ)\mid\\
& \alpha_Q=\pi_{TQ}\circ\nu_{\alpha_Q},~\nu_{\alpha_Q}(\tilde\tau)\in T_{\alpha_Q(\tilde\tau)}\partial Q,\nonumber\\
& \nu_{\alpha_Q}\text{ is piecewise }C^2\text{ and has only one singularity at $\tilde\tau$}\Big\},\nonumber
\end{align}
where $\pi_{TQ}:TQ\to Q$ is the natural projection. In order to incorporate the constraint distribution, we define the following subspace at each $\alpha_Q\in\mathcal Q(\tilde\tau)$,
\begin{equation*}
\Delta_{\mathcal Q(\tilde\tau)}(\alpha_Q)=\left\{\nu_{\alpha_Q}\in T_{\alpha_Q}\mathcal Q(\tilde\tau)\mid\nu_{\alpha_Q}:[\tau_0,\tau_1]\to\Delta_Q\right\}.
\end{equation*}
As usual, we denote $T\mathcal Q(\tilde\tau)=\bigsqcup_{\alpha_Q\in\mathcal Q(\tilde\tau)}T_{\alpha_Q}\mathcal Q(\tilde\tau)$ and $\Delta_{\mathcal Q(\tilde\tau)}=\bigsqcup_{\alpha_Q\in\mathcal Q(\tilde\tau)}\Delta_{\mathcal Q(\tilde\tau)}(\alpha_Q)$.

Let $T_{\alpha_Q}'\mathcal Q(\tilde\tau)=\{\phi_{\alpha_Q}:T_{\alpha_Q}\mathcal Q(\tilde\tau)\to\mathbb R\mid\phi_{\alpha_Q}\text{ is linear}\newline \text{ and continuous}\}$ be the topological dual of $T_{\alpha_Q}\mathcal Q(\tilde\tau)$. Since $\mathcal Q(\tilde\tau)$ is an infinite dimensional manifold, its topological cotangent bundle is too large to formulate mechanics. For that reason, we will restrict ourselves to the vector subbundle where the Legendre transform of the Lagrangian lie, i.e., we consider a vector subbundle $T^\star\mathcal Q(\tilde\tau)\subset T'\mathcal Q(\tilde\tau)$ such that $\mathbb FL\circ\nu_{\alpha_Q}\in T^\star\mathcal Q(\tilde\tau))$ for each $\nu_{\alpha_Q}\in T_{\alpha_Q}\mathcal Q(\tilde\tau)$, where $\mathbb FL:TQ\to T^*Q$ is the Legendre transform of $L$.

\begin{lemma}
For each $\alpha_Q\in\mathcal Q(\tilde\tau)$, the vector space
\begin{align}\label{eq:T*mathcalQ}
& T_{\alpha_Q}^\star\mathcal Q(\tilde\tau)=\big\{\pi_{\alpha_Q}\in C^0([\tau_0,\tau_1],T^*Q)\mid\\ &\alpha_Q=\pi_{T^*Q}\circ\pi_{\alpha_Q},~\pi_{\alpha_Q}(\tilde\tau)\in T_{\alpha_Q(\tau)}^*\partial Q,\nonumber\\
& \pi_{\alpha_Q}\text{ is piecewise }C^2\text{ and has only one singularity at $\tilde\tau$}\big\},\nonumber
\end{align}
where $\pi_{T^*Q}:T^*Q\to Q$ is the natural projection, is a vector subspace of the topological dual of $T_{\alpha_Q}\mathcal Q(\tilde\tau)$ by means of the following $L^2$-dual pairing:
\begin{equation*}
\langle\pi_{\alpha_Q},\nu_{\alpha_Q}\rangle=\int_{\tau_0}^{\tau_1}\pi_{\alpha_Q}(\tau)\cdot\nu_{\alpha_Q}(\tau)\,d\tau,
\end{equation*}
where $\cdot$ represents the pairing between $T^*Q$ and $TQ$. Furthermore, this pairing is nondegenerate.
\end{lemma}

Observe that, in general, $$\left\{\mathbb FL\circ\nu_{\alpha_Q}\in T_{\alpha_Q}^\star\mathcal Q(\tilde\tau)\mid\nu_{\alpha_Q}\in T_{\alpha_Q}\mathcal Q(\tilde\tau)\right\}\subsetneq T_{\alpha_Q}^\star\mathcal Q(\tilde\tau),$$ as the Lagrangian is possibly degenerate. As a straightforward consequence of the previous lemma, the vector bundle
\begin{equation*}
T^\star\mathcal Q(\tilde\tau)=\bigsqcup_{\alpha_Q\in\mathcal Q(\tilde\tau)}T_{\alpha_Q}^\star\mathcal Q(\tilde\tau)\to\mathcal Q(\tilde\tau),\quad\pi_{\alpha_Q}\mapsto\alpha_Q,
\end{equation*}
is a vector subbundle of the topological cotangent bundle of $\mathcal Q(\tilde\tau)$. 

In the same vein, for each $\nu_{\alpha_Q}\in T_{\alpha_Q}\mathcal Q(\tilde\tau)$ and $\pi_{\alpha_Q}\in T_{\alpha_Q}^\star\mathcal Q(\tilde\tau)$, the iterated bundles are given by
\begin{align*}
& T_{\nu_{\alpha_Q}}(T\mathcal Q(\tilde\tau))=\Big\{\delta\nu_{\alpha_Q}\in C^0([\tau_0,\tau_1],T(TQ))\mid\\
& \nu_{\alpha_Q}=\pi_{T(TQ)}\circ\delta\nu_{\alpha_Q},~\delta\nu_{\alpha_Q}(\tilde\tau)\in T_{\nu_{\alpha_Q}(\tilde\tau)}(T\partial Q),\\
& \delta\nu_{\alpha_Q}\text{ is piecewise $C^2$ and has only one singularity at }\tilde\tau\Big\},
\end{align*}
where $\pi_{T(TQ)}:T(TQ)\to TQ$ is the natural projection, and
\begin{align*}
& T_{\pi_{\alpha_Q}}(T^\star\mathcal Q(\tilde\tau))=\Big\{\delta\pi_{\alpha_Q}\in C^0([\tau_0,\tau_1],T(T^*Q))\mid\\
& \pi_{\alpha_Q}=\pi_{T(T^*Q)}\circ\delta\pi_{\alpha_Q},~\delta\pi_{\alpha_Q}(\tilde\tau)\in T_{\pi_{\alpha_Q}(\tilde\tau)}(T^*\partial Q),\\
& \delta\pi_{\alpha_Q}\text{ is piecewise $C^2$ and has only one singularity at }\tilde\tau\Big\},
\end{align*}
where $\pi_{T(T^*Q)}:T(T^*Q)\to T^*Q$ is the natural projection. In particular, we consider the constrained iterated bundle,
\begin{align}\label{eq:DeltaTQ}
\Delta_{T\mathcal Q(\tilde\tau)}(\nu_{\alpha_Q})=&\left\{\delta\nu_{\alpha_Q}\in T_{\nu_{\alpha_Q}}(T\mathcal Q(\tilde\tau))\mid \right.\\ & \left.d\pi_{TQ}\circ\delta\nu_{\alpha_Q}\in C^0([\tau_0,\tau_1],\Delta_Q)\right\}\nonumber.
\end{align}

\section{Nonholonomic implicit Lagrangian mechanics with collisions}\label{sec4}
Given a path $\alpha=(\alpha_T,\alpha_Q)\in\Omega(Q,\tilde\tau)$, the \emph{associated curve} is defined as
\begin{equation}\label{eq:associatedcurve}
q_\alpha:[t_0,t_1]\to Q,\quad t\mapsto q_\alpha(t)=\left(\alpha_Q\circ\alpha_T^{-1}\right)(t).
\end{equation}
Similarly, given $\nu_{\alpha_Q}\in T_{\alpha_Q}\mathcal Q(\tilde\tau)$ and $\pi_{\alpha_Q}\in T_{\alpha_Q}'\mathcal Q(\tilde\tau)$, we set
\begin{equation*}
\begin{array}{ll}
v_\alpha: [t_0,t_1]\to TQ, \quad & t\mapsto v_\alpha(t)=\left(\nu_{\alpha_Q}\circ\alpha_T^{-1}\right)(t),\\
p_\alpha: [t_0,t_1]\to T^*Q, & t\mapsto p_\alpha(t)=\left(\pi_{\alpha_Q}\circ\alpha_T^{-1}\right)(t).
\end{array}
\end{equation*}
It is clear that $\pi_{TQ}\circ v_\alpha=\pi_{T^*Q}\circ p_\alpha=q_\alpha$.

By regarding $\Omega(Q,\tilde\tau)$ as a trivial vector bundle over $\mathcal Q(\tilde\tau)$ with the projection onto the second factor, the \emph{Lagrange--d'Alembert--Pontryagin action functional},
\begin{equation*}
\mathbb S:\Omega(Q,\tilde\tau)\times_{\mathcal Q(\tilde\tau)}\left(T\mathcal Q(\tilde\tau)\oplus T^\star\mathcal Q(\tilde\tau)\right)\to\mathbb R,
\end{equation*}
where $\times_{\mathcal Q(\tilde\tau)}$ denotes the fibered product over $\mathcal Q(\tilde\tau)$, is defined as
\begin{align*}
&\mathbb{S}\left(\alpha,\nu_{\alpha_Q},\pi_{\alpha_Q}\right)  =\int_{t_0}^{t_1}\left(L(v_\alpha(t))+p_\alpha(t)\cdot\left(\dot q_\alpha(t)-v_\alpha(t)\right)\right)dt\\
& =\int_{\tau_0}^{\tau_1}\left(L\left(\nu_{\alpha_Q}(\tau)\right)+\pi_{\alpha_Q}(\tau)\cdot\left(\frac{\alpha_Q'(\tau)}{\alpha_T'(\tau)}-\nu_{\alpha_Q}(\tau)\right)\right)\alpha_T'(\tau)\,d\tau.
\end{align*}
The equality between the first and the second expressions can be easily checked by considering the change of variable $t=\alpha_T(\tau)$. By recalling that the \emph{energy} of the system, $E:TQ\oplus T^*Q\to\mathbb R$, is given by
\begin{equation}\label{eq:energy}
E(v_q,p_q)=p_q\cdot v_q-L(v_q),\quad(v_q,p_q)\in TQ\oplus T^*Q,
\end{equation}
the action functional may be rewritten as
\begin{align*}
&\mathbb S\left(\alpha,\nu_{\alpha_Q},\pi_{\alpha_Q}\right)  =\int_{t_0}^{t_1}\left(p_\alpha(t)\cdot\dot q_\alpha(t)-E(v_\alpha(t),p_\alpha(t)\right)dt\\
& =\int_{\tau_0}^{\tau_1}\left(\pi_{\alpha_Q}(\tau)\cdot\frac{\alpha_Q'(\tau)}{\alpha_T'(\tau)}-E\left(\nu_{\alpha_Q}(\tau),\pi_{\alpha_Q}(\tau)\right)\right)\alpha_T'(\tau)\,d\tau.
\end{align*}

\begin{definition}[Hamilton--d'Alembert--Pontryagin  principle]\label{def:HdAPprinciple}
A path
\begin{equation*}
\texttt{c}=((\alpha_T,\alpha_Q),\nu_{\alpha_Q},\pi_{\alpha_Q})\in\Omega(Q,\tilde\tau)\times_{\mathcal Q(\tilde\tau)}\left(\Delta_{\mathcal Q(\tilde\tau)}\oplus T^\star\mathcal Q(\tilde\tau)\right)
\end{equation*}
is \emph{stationary} (or \emph{critical}) for the action functional $\mathbb S$ if it satisfies
$$d\mathbb S(\texttt{c})(\delta\texttt{c})=0,$$ for every variation $\delta\texttt{c}=\left((\delta\alpha_T,\delta\alpha_Q),\delta\nu_{\alpha_Q},\delta\pi_{\alpha_Q}\right)\in T_\alpha\Omega(Q,\tilde\tau)\times\Delta_{T\mathcal Q(\tilde\tau)}(\nu_{\alpha_Q})\times T_{\pi_{\alpha_Q}}(T^\star\mathcal Q(\tilde\tau))$ such that $\delta\alpha_T(\tau_0)=\delta\alpha_T(\tau_1)=0$, $\delta\alpha_Q(\tau_0)=\delta\alpha_Q(\tau_1)=0$ and
\begin{equation}\label{eq:projectionvariation}
d\pi_{TQ}\circ\delta\nu_{\alpha_Q}=d\pi_{T^*Q}\circ\delta\pi_{\alpha_Q}=\delta\alpha_Q.
\end{equation}
\end{definition}

\begin{theorem}\label{theorem:implicitELequations}
A (local) curve
\begin{align*}
&(\alpha,\nu_{\alpha_Q},\pi_{\alpha_Q})\simeq\\&(\alpha_T,\alpha_Q,\nu_Q,\pi_Q)\in\Omega(Q,\tilde\tau)\times_{\mathcal Q(\tilde\tau)}\left(\Delta_{\mathcal Q(\tilde\tau)}\oplus T^*\mathcal Q(\tilde\tau)\right)    \end{align*}
is critical for the action functional $\mathbb S$ if and only if it satisfies the \emph{implicit Euler--Lagrange equations},
\begin{equation*}
\left\{\begin{array}{ll}
\displaystyle\pi_Q'-\alpha_T'\frac{\partial L}{\partial q}(\alpha_Q,\nu_Q)\in\Delta_Q^\circ(\alpha_Q),\, & \displaystyle E'(\alpha_Q,\nu_Q,\pi_Q)=0,\\
\displaystyle\pi_Q=\frac{\partial L}{\partial v}(\alpha_Q,\nu_Q), & \displaystyle\nu_Q=\frac{\alpha_Q'}{\alpha_T'}\in\Delta_Q(\alpha_Q),
\end{array}\right.
\end{equation*}
on $[\tau_0,\tilde\tau)\cup(\tilde\tau,\tau_1]$, together with the conditions for the \emph{elastic impact},
\begin{align*}
& \pi_Q(\tilde\tau^+)-\pi_Q(\tilde\tau^-)\in(T\partial Q\cap\Delta_Q)^\circ=(T\partial Q)^\circ+\Delta_Q^\circ,\\
& E(\alpha_Q(\tilde\tau^-),\nu_Q(\tilde\tau^-),\pi_Q(\tilde\tau^-))=E(\alpha_Q(\tilde\tau^+),\nu_Q(\tilde\tau^+),\pi_Q(\tilde\tau^+)),
\end{align*}
where the annihilitaros are with respect to $TQ$.
\end{theorem}

\textit{Proof:} Let $TQ\simeq Q\times V$ be a trivialization of the tangent bundle of $Q$, and consider the induced trivializations of the cotangent bundle of $Q$, $T^*Q\simeq Q\times V^*$, as well as of the iterated bundles $T(TQ)\simeq Q\times V\times V\times V$ and $T(T^*Q)\simeq Q\times V^*\times V\times V^*$. Locally, we may write $\alpha'\simeq(\alpha_T,\alpha_Q,\alpha_T',\alpha_Q')$, $\nu_{\alpha_Q}\simeq(\alpha_Q,\nu_Q)$ and $\pi_{\alpha_Q}\simeq(\alpha_Q,\pi_Q)$ for some $\alpha_Q',\nu_Q:[\tau_0,\tau_1]\to V$ and $\pi_Q:[\tau_0,\tau_1]\to V^*$. Moreover, the variations locally read $\delta\alpha\simeq(\alpha_T,\alpha_Q,\delta\alpha_T,\delta\alpha_Q)$, $\delta\nu_{\alpha_Q}\simeq(\alpha_Q,\nu_Q,\beta_Q,\delta\nu_Q)$ and $\delta\pi_{\alpha_Q}\simeq(\alpha_Q,\pi_Q,\gamma_Q,\delta\pi_Q)$ for some $\delta\alpha_Q,\beta_Q,\gamma_Q,\delta\nu_Q:[\tau_0,\tau_1]\to V$, $\delta\pi_Q:[\tau_0,\tau_1]\to V^*$. In fact, equation \eqref{eq:projectionvariation} ensures that $\delta\alpha_Q=\beta_Q=\gamma_Q$. Moreover, by locally regarding $\Delta_Q(q)\subset V$ for each $q\in Q$, the conditions $\nu_{\alpha_Q}\in\Delta_{\mathcal Q(\tilde\tau)}$ and $\delta\nu_{\alpha_Q}\in\Delta_{T\mathcal Q(\tilde\tau)}(\nu_{\alpha_Q})$ read $\nu_Q(\tau)\in\Delta_Q(\alpha_Q(\tau))$ and $\delta\alpha_Q(\tau)\in\Delta_Q(\alpha_Q(\tau))$ for each $\tau\in[\tau_0,\tau_1]$, respectively. At last, the condition $\delta\nu_{\alpha_Q}(\tilde\tau)\in T_{\nu_{\alpha_Q}(\tilde\tau)}(T\partial Q)$ yields the local condition $\delta\alpha_Q(\tilde\tau)\in W$, where $W\subset V$ is a subspace of co-dimension one such that $T\partial Q\simeq\partial Q\times W$.

As a result, the variation of the action functional reads
\begin{align*}
& d\mathbb S(\alpha,\nu_{\alpha_Q},\pi_{\alpha_Q})\left(\delta\alpha,\delta\nu_{\alpha_Q},\delta\pi_{\alpha_Q}\right)\simeq\\ 
& d\mathbb S(\alpha_T,\alpha_Q,\nu_Q,\pi_Q)(\delta\alpha_T,\delta\alpha_Q,\delta\nu_Q,\delta\pi_Q)=\\
& \int_{\tau_0}^{\tau_1}\left(\frac{\partial L}{\partial q}\cdot\delta\alpha_Q+\frac{\partial L}{\partial v}\cdot\delta\nu_Q+\delta\pi_Q\cdot\left(\frac{\alpha_Q'}{\alpha_T'}-\nu_Q\right)\right.\\ 
& \hspace{8mm}\left.+\pi_Q\cdot\left(\frac{\delta\alpha_Q'}{\alpha_T'}-\frac{\alpha_Q'\delta\alpha_T'}{(\alpha_T')^2}-\delta\nu_Q\right)\right)\alpha_T'\,d\tau\\
& \hspace{8mm}+\int_{\tau_0}^{\tau_1}\left(L+\pi_Q\cdot\left(\frac{\alpha_Q'}{\alpha_T'}-\nu_Q\right)\right)\delta\alpha_T'\,d\tau,
\end{align*}
where the Lagrangian, as well as its partial derivatives, are evaluated at $\left(\alpha_Q,\nu_Q\right)$. After splitting the integration domain, $[\tau_0,\tau_1]-\{\tilde\tau\}=[\tau_0,\tilde\tau)\cup(\tilde\tau,\tau_1]$, as well as integrating by parts on each sub-interval, we may rewrite the previous expression as
\begin{align*}
& d\mathbb S(\alpha,\nu_{\alpha_Q},\pi_{\alpha_Q})\left(\delta\alpha,\delta\nu_{\alpha_Q},\delta\pi_{\alpha_Q}\right)\simeq\\
& \int_{\tau_0}^{\tilde\tau}\left(\left(\alpha_T'\frac{\partial L}{\partial q}-\pi_Q'\right)\cdot\delta\alpha_Q-\frac{d}{d\tau}(L-\pi_Q\cdot\nu_Q)\delta\alpha_T\right.\\
& \hspace{8mm}\left.+\alpha_T'\left(\frac{\partial L}{\partial v}-\pi_Q\right)\cdot\delta\nu_Q+\delta\pi_Q\cdot(\alpha_Q'-\alpha_T'\nu_Q)\right)d\tau\\
& +\int_{\tilde\tau}^{\tau_1}\left(\left(\alpha_T'\frac{\partial L}{\partial q}-\pi_Q'\right)\cdot\delta\alpha_Q-\frac{d}{d\tau}(L-\pi_Q\cdot\nu_Q)\delta\alpha_T\right.\\
& \hspace{8mm}\left.+\alpha_T'\left(\frac{\partial L}{\partial v}-\pi_Q\right)\cdot\delta\nu_Q+\delta\pi_Q\cdot(\alpha_Q'-\alpha_T'\nu_Q)\right)d\tau\\
& +\Big[\pi_Q\cdot\delta\alpha_Q+(L-\pi_Q\cdot\nu_Q)\delta\alpha_T\Big]_{\tau=\tau_0}^{\tau=\tilde\tau^-}\\
& +\Big[\pi_Q\cdot\delta\alpha_Q+(L-\pi_Q\cdot\nu_Q)\delta\alpha_T\Big]_{\tau=\tilde\tau^+}^{\tau=\tau_1}.
\end{align*}
Since the previous expression vanishes for free variations $(\delta\alpha,\delta\nu_Q,\delta\pi_Q)$ such that $\delta\alpha_Q\in\Delta_Q(\alpha_Q)$, $\delta\alpha_T(\tau_0)=\delta\alpha_T(\tau_1)=0$ and $\delta\alpha_Q(\tau_0)=\delta\alpha_Q(\tau_1)=0$, we obtain the desired equations and impact conditions.
\hfill$\square$

By using the change of variable $t=\alpha_T(\tau)$, we have $\dot q_\alpha=\alpha_Q'/\alpha_T'$ and $\dot p_\alpha=\pi_Q'/\alpha_T'$. Then, the implicit Euler--Lagrange equations for a (local) curve
\begin{equation*}
(v_\alpha,p_\alpha)\simeq(q_\alpha,v,p):[t_0,t_1]\to TQ\oplus T^*Q    
\end{equation*}
take the form
\begin{equation}\label{eq:ELeqsimplicit}\left\{\begin{array}{ll}
\displaystyle\dot p-\frac{\partial L}{\partial q}(q_\alpha,v)\in\Delta_Q^\circ(q_\alpha),\qquad & \displaystyle \dot E(q_\alpha,v,p)=0,\vspace{0.1cm}\\
\displaystyle p=\frac{\partial L}{\partial v}(q_\alpha,v), & \displaystyle v=\dot q_\alpha\in\Delta_Q(q_\alpha),
\end{array}\right.
\end{equation}
on $\left[t_0,\tilde t\right)\cup\left(\tilde t,t_1\right]$. Similarly, the conditions for the elastic impact read
\begin{align}\label{eq:impactL}
& p\left(\tilde t^+\right)-p\left(\tilde t^-\right)\in(T\partial Q\cap\Delta_Q)^\circ=(T\partial Q)^\circ+\Delta_Q^\circ,\\\nonumber
& E\left(q_\alpha\left(\tilde t^-\right),v\left(\tilde t^-\right),p\left(\tilde t^-\right)\right)=E\left(q_\alpha\left(\tilde t^+\right),v\left(\tilde t^+\right),p\left(\tilde t^+\right)\right),\\\label{eq:impactDelta}
& v\left(\tilde t^+\right)=\dot q_\alpha\left(\tilde t^+\right)\in\Delta_Q.
\end{align}

\textbf{Energy balance:} It may be shown that the conservation of the energy along the solutions, $\dot E(q_\alpha,v,p)=0$, is redundant, as it may be obtained from the remaining equations.

For unconstrained systems, i.e., $\Delta_Q=TQ$, the Hamilton--d'Alembert--Pontryagin principle reduces to the Hamilton--Pontryagin principle, and the implicit Euler--Lagrange equations of motion read as
\begin{equation*}
\dot p=\frac{\partial L}{\partial q}(q_\alpha,v),\qquad p=\frac{\partial L}{\partial v}(q_\alpha,v),\qquad v=\dot q_\alpha.
\end{equation*}

\section{Nonholonomic implicit Hamiltonian mechanics with collisions}\label{sec5}

The results in the previous section may be obtained in the Hamiltonian side as well. Namely, given a (possibly degenerate) Hamiltonian, $H:T^*Q\to\mathbb R$, the \emph{Hamilton--d'Alembert--Pontryagin action functional},
\begin{equation*}
\mathfrak S:\Omega(Q,\tilde\tau)\times_{\mathcal Q(\tilde\tau)}T^\star\mathcal Q(\tilde\tau)\to\mathbb R,
\end{equation*}
is defined as
\begin{align*}
&\mathfrak S\left(\alpha,\pi_{\alpha_Q}\right)  =\int_{t_0}^{t_1}\big(p_\alpha(t)\cdot\dot q_\alpha(t)-H(p_\alpha(t))\big)\,dt\\
& =\int_{\tau_0}^{\tau_1}\left(\pi_{\alpha_Q}(\tau)\cdot\frac{\alpha_Q'(\tau)}{\alpha_T'(\tau)}-H\left(\pi_{\alpha_Q}(\tau)\right)\right)\alpha_T'(\tau)\,d\tau.
\end{align*}
Recall that from this point of view the \emph{energy} of the system is simply given by the Hamiltonian.

\begin{definition}[Variational principle in the phase space]\label{def:HdAPprincipleH}
A path
\begin{equation*}
\texttt{c}=((\alpha_T,\alpha_Q),\pi_{\alpha_Q})\in\Omega(Q,\tilde\tau)\times_{\mathcal Q(\tilde\tau)}T^\star\mathcal Q(\tilde\tau)
\end{equation*}
such that $\alpha_Q'\in\Delta_{\mathcal Q(\tilde\tau)}(\alpha_Q)$ is \emph{stationary} (or \emph{critical}) for the action functional $\mathfrak S$ if it satisfies $d\mathfrak S(\texttt{c})(\delta\texttt{c})=0$
for every variation $\delta\texttt{c}=\left((\delta\alpha_T,\delta\alpha_Q),\delta\pi_{\alpha_Q}\right)\in T_\alpha\Omega(Q,\tilde\tau)\times T_{\pi_{\alpha_Q}}(T^\star\mathcal Q(\tilde\tau))$ such that $\delta\alpha_T(\tau_0)=\delta\alpha_T(\tau_1)=0$, $\delta\alpha_Q(\tau_0)=\delta\alpha_Q(\tau_1)=0$ and
\begin{equation}\label{eq:projectionvariationH}
d\pi_{T^*Q}\circ\delta\pi_{\alpha_Q}=\delta\alpha_Q\in\Delta_{\mathcal Q(\tilde\tau)}(\alpha_Q).
\end{equation}
\end{definition}

\begin{theorem}\label{theorem:implicitHequations}
A (local) curve
\begin{equation*}
(\alpha,\pi_{\alpha_Q})\simeq(\alpha_T,\alpha_Q,\pi_Q)\in\Omega(Q,\tilde\tau)\times_{\mathcal Q(\tilde\tau)}T^*\mathcal Q(\tilde\tau)    
\end{equation*}
such that $\alpha_Q'\in\Delta_{\mathcal Q(\tilde\tau)}(\alpha_Q)$ is critical for the action functional $\mathfrak S$ if and only if it satisfies the \emph{implicit Hamilton equations},
\begin{align*}
& \pi_Q'+\alpha_T'\frac{\partial H}{\partial q}(\alpha_Q,\pi_Q)\in\Delta_Q^\circ(\alpha_Q),\qquad H'(\alpha_Q,\pi_Q)=0,\\
& \frac{\alpha_Q'}{\alpha_T'}=\frac{\partial H}{\partial p}(\alpha_Q,\pi_Q)\in\Delta_Q(\alpha_Q),
\end{align*}
on $[\tau_0,\tilde\tau)\cup(\tilde\tau,\tau_1]$, together with the conditions for the \emph{elastic impact},
\begin{equation*}
\left\{\begin{array}{l}
\displaystyle\pi_Q(\tilde\tau^+)-\pi_Q(\tilde\tau^+)\in(T\partial Q\cap\Delta_Q)^\circ=(T\partial Q)^\circ+\Delta_Q^\circ,\vspace{2mm}\\
\displaystyle H(\alpha_Q(\tilde\tau^-),\pi_Q(\tilde\tau^-))=H(\alpha_Q(\tilde\tau^+),\pi_Q(\tilde\tau^+)).
\end{array}\right.
\end{equation*}
\end{theorem}

\textit{Proof:} Let $TQ\simeq Q\times V$ be a trivialization of the tangent bundle of $Q$, and consider the induced trivializations of the cotangent bundle of $Q$, $T^*Q\simeq Q\times V^*$, as well as of the iterated bundle $T(T^*Q)\simeq Q\times V^*\times V\times V^*$. Locally, we may write $\alpha'\simeq(\alpha_T,\alpha_Q,\alpha_T',\alpha_Q')$ and $\pi_{\alpha_Q}\simeq(\alpha_Q,\pi_Q)$ for some $\alpha_Q':[\tau_0,\tau_1]\to V$ and $\pi_Q:[\tau_0,\tau_1]\to V^*$. Moreover, the variations locally read $\delta\alpha\simeq(\alpha_T,\alpha_Q,\delta\alpha_T,\delta\alpha_Q)$ and $\delta\pi_{\alpha_Q}\simeq(\alpha_Q,\pi_Q,\gamma_Q,\delta\pi_Q)$ for some $\delta\alpha_Q,\gamma_Q:[\tau_0,\tau_1]\to V$ and $\delta\pi_Q:[\tau_0,\tau_1]\to V^*$. In fact, equation \eqref{eq:projectionvariationH} ensures that $\delta\alpha_Q=\gamma_Q\in\Delta_Q(\alpha_Q)$. Moreover, we have $\delta\alpha_Q(\tilde\tau)\in W$, where $W\subset V$ is a subspace of co-dimension one such that $T\partial Q\simeq\partial Q\times W$.

As a result, the variation of the action functional reads
\begin{align*}
& d\mathfrak S(\alpha,\pi_{\alpha_Q})\left(\delta\alpha,\delta\pi_{\alpha_Q}\right)\simeq\\ 
& d\mathfrak S(\alpha_T,\alpha_Q,\pi_Q)(\delta\alpha_T,\delta\alpha_Q,\delta\pi_Q)=\\
& \int_{\tau_0}^{\tau_1}\bigg(\delta\pi_Q\cdot\frac{\alpha_Q'}{\alpha_T'}+\pi_Q\cdot\left(\frac{\delta\alpha_Q'}{\alpha_T'}-\frac{\alpha_Q'\delta\alpha_T'}{(\alpha_T')^2}\right)\\
& \hspace{8mm}-\frac{\partial H}{\partial q}\cdot\delta\alpha_Q-\frac{\partial H}{\partial p}\cdot\delta\pi_Q\bigg)\,\alpha_T'\,d\tau\\
& +\int_{\tau_0}^{\tau_1}\left(\pi_Q\cdot\frac{\alpha_Q'}{\alpha_T'}-H\right)\delta\alpha_T'\,d\tau,
\end{align*}
where the Hamiltonian, as well as its partial derivatives, are evaluated at $\left(\alpha_Q,\pi_Q\right)$. After splitting the integration domain, $[\tau_0,\tau_1]-\{\tilde\tau\}=[\tau_0,\tilde\tau)\cup(\tilde\tau,\tau_1]$, as well as integrating by parts on each sub-interval, we may rewrite the previous expression as
\begin{align*}
& d\mathfrak S(\alpha,\pi_{\alpha_Q})\left(\delta\alpha,\delta\pi_{\alpha_Q}\right)\simeq\\
& \int_{\tau_0}^{\tilde\tau}\bigg(\left(-\pi_Q'-\alpha_T'\frac{\partial H}{\partial q}\right)\cdot\delta\alpha_Q+H'\delta\alpha_T\\
& \hspace{8mm}+\delta\pi_Q\cdot\left(\alpha_Q'-\alpha_T'\frac{\partial H}{\partial p}\right)\bigg)\,d\tau\\
& +\int_{\tilde\tau}^{\tau_1}\bigg(\left(-\pi_Q'-\alpha_T'\frac{\partial H}{\partial q}\right)\cdot\delta\alpha_Q+H'\delta\alpha_T\\
& \hspace{8mm}+\delta\pi_Q\cdot\left(\alpha_Q'-\alpha_T'\frac{\partial H}{\partial p}\right)\bigg)\,d\tau\\
& +\Big[\pi_Q\cdot\delta\alpha_Q-H\delta\alpha_T\Big]_{\tau=\tau_0}^{\tau=\tilde\tau^-}+\Big[\pi_Q\cdot\delta\alpha_Q-H\delta\alpha_T\Big]_{\tau=\tilde\tau^+}^{\tau=\tau_1}.
\end{align*}
Since the previous expression vanishes for free variations $(\delta\alpha,\delta\pi_Q)$ such that $\delta\alpha_Q\in\Delta_Q(\alpha_Q)$, $\delta\alpha_T(\tau_0)=\delta\alpha_T(\tau_1)=0$ and $\delta\alpha_Q(\tau_0)=\delta\alpha_Q(\tau_1)=0$, we obtain the desired equations and impact conditions.
\hfill$\square$

As for the Lagrangian equations, by means of the change of variable $t=\alpha_T(\tau)$, the implicit Hamilton equations for a (local) curve
\begin{equation*}
p_\alpha\simeq(q_\alpha,p):[t_0,t_1]\to T^*Q    
\end{equation*}
take the form
\begin{align*}
& \dot p+\frac{\partial H}{\partial q}(q_\alpha,p)\in\Delta_Q^\circ(q_\alpha),\qquad \dot H(q_\alpha,p)=0,\\
& \dot q_\alpha=\frac{\partial H}{\partial p}(q_\alpha,p)\in\Delta_Q(q_\alpha),
\end{align*}
on $\left[t_0,\tilde t\right)\cup\left(\tilde t,t_1\right]$. Similarly, the conditions for the elastic impact read
\begin{align*}
& p\left(\tilde t^+\right)-p\left(\tilde t^-\right)\in(T\partial Q\cap\Delta_Q)^\circ=(T\partial Q)^\circ+\Delta_Q^\circ,\\
& H\left(q_\alpha\left(\tilde t^-\right),p\left(\tilde t^-\right)\right)=H\left(q_\alpha\left(\tilde t^+\right),p\left(\tilde t^+\right)\right)\\
& \dot q_\alpha\left(\tilde t^+\right)\in\Delta_Q.
\end{align*}

\textbf{Energy balance:} It may be shown that the conservation of the energy along the solutions, $\dot H(q_\alpha,p)=0$, is redundant, as it may be obtained from the remaining equations.

For unconstrained systems, i.e., $\Delta_Q=TQ$, the Hamilton--d'Alembert--Pontryagin principle in the phase space reduces to the Hamilton--Pontryagin principle in the phase space, and the implicit Hamilton equations of motion read as
\begin{equation*}
\dot p=-\frac{\partial H}{\partial q}(q_\alpha,p),\qquad \dot q_\alpha=\frac{\partial H}{\partial p}(q_\alpha,p).
\end{equation*}

\textbf{Hiperregular Lagrangians:} When $L:TQ\to\mathbb R$ is a hyperregular Lagrangian, i.e., when the Legendre transform $\mathbb FL:TQ\to T^*Q$ is an isomorphism, then both the Lagrangian and the Hamiltonian approaches are equivalent. Namely, the Lagrangian $L$ induces the Hamiltomian $H:T^*Q\to\mathbb R$ given by
\begin{equation*}
H(p_q)=E\left((\mathbb FL)^{-1}(p_q),p_q\right),\quad p_q\in T^*Q.
\end{equation*}
By recalling the local expression of the Legendre transform,
\begin{equation*}
\mathbb FL(q,v)=\left(q,\frac{\partial L}{\partial v}(q,v)\right),
\end{equation*}
it is easy to check that the implicit Euler--Lagrange equations together with the conditions for the elastic impact hold for a (local) curve $(v_\alpha,p_\alpha):[t_0,t_1]\to TQ\oplus T^*Q$ if and only if $p_\alpha=\mathbb FL(v_\alpha)$ and the implicit Hamilton equations together with the conditions for the elastic impact hold for the (local) curve $p_\alpha=\mathbb FL(v_\alpha):[t_0,t_q]\to T^*Q$.

\section{Rolling disk hitting a wall}\label{sec6}

Let us consider a disk rolling without slipping, as in \cite[Section 7.1]{YoMa2006a}. However, here we assume that there is a wall that the disk may hit (see \cite{simoes2023hamel}). The configuration space is thus given by
\begin{equation*}
Q=\{(x,y,\theta,\varphi)\in\mathbb R^2\times\mathbb S^1\times\mathbb S^1\mid y+ R\sin\varphi\leq10\},
\end{equation*}
where $(x,y)$ denotes the contact point of the disk with the ground, $\theta$ denotes the angle of rotation and $\varphi$ denotes the heading angle of the disk with respect to the $x$-axis. The Lagrangian $L:TQ\to\mathbb R$ is given by
\begin{equation*}
L(x,y,\theta,\varphi;v_x,v_y,v_\theta,v_\varphi)=\frac{1}{2}m\left(v_x^2+v_y^2\right)+\frac{1}{2}\left(I\,v_\theta^2+J\,v_\varphi^2\right),
\end{equation*}
where $m,I,J\in\mathbb R^+$ are the mass and the moments of inertia of the disk, respectively. For each $(v_q,p_q)=(x,y,\theta,\varphi;v_x,v_y,v_\theta,v_\varphi;p_x,p_y,p_\theta,p_\varphi)\in TQ\oplus T^*Q$, the energy reads
\begin{align*}
 E(v_q,p_q)=& p_x\,v_x+p_y\,v_y+p_\theta\,v_\theta+p_\varphi\,v_\varphi\\
& -\frac{1}{2}m\left(v_x^2+v_y^2\right)-\frac{1}{2}\left(I\,v_\theta^2+Jv_\varphi^2\right).
\end{align*}

\textbf{Non-holonomic constraint:} The non-slipping condition reads $v_x=R\, v_\theta\cos\varphi,$ $v_y=R\,v_\theta\sin\varphi$, where $R\in\mathbb R^+$ is the radius of the disk, thus yielding following non-holonomic constraint:
\begin{equation*}
\Delta_Q=\operatorname{span}\{\partial_\theta+R\cos\varphi\,\partial_x+R\sin\varphi\,\partial_y,\partial_\varphi\}.
\end{equation*}
The annihilator is easily seen to be
\begin{equation*}
\Delta_Q^\circ=\operatorname{span}\{dx-R\cos\varphi\,d\theta,dy-R\sin\varphi\,d\theta\}.
\end{equation*}
On the other hand, the boundary of the configuration manifold is given by
\begin{equation*}
\partial Q=\{(x,y,\theta,\varphi)\in\mathbb R^2\times\mathbb S^1\times\mathbb S^1\mid y+ R\sin\varphi=10\},
\end{equation*}
whose tangent bundle reads
\begin{equation*}
T\partial Q=\operatorname{span}\{\partial_x,\partial_\theta,\partial_\varphi-R\cos\varphi\,\partial_y\}.
\end{equation*}
Therefore, its annihilator reads
\begin{equation*}
(T\partial Q)^\circ=\operatorname{span}\{dy+R\cos\varphi\,d\varphi\}.
\end{equation*}

\textbf{Dynamical equations:} The implicit Euler--Lagrange equations with collisions for a curve
\begin{equation*}
(x,y,\theta,\varphi;v_x,v_y,v_\theta,v_\varphi;p_x,p_y,p_\theta,p_\varphi):[t_0,t_1]\to TQ\oplus T^*Q    
\end{equation*}
given in \eqref{eq:ELeqsimplicit} read
\begin{equation*}
\left\{\begin{array}{ll}
R\,\dot p_x\cos\varphi+R\,\dot p_y\sin\varphi+\dot p_\theta=0,\qquad & \dot p_\varphi=0,\\
v_x=R\,v_\theta\cos\varphi,\qquad & v_y=R\,v_\theta\sin\varphi,\\
p_x=m\,v_x,\qquad & p_y=m\,v_y,\\
p_\theta=I\,v_\theta,\qquad & p_\varphi=J\,v_\varphi,\\
v_x=\dot x,\qquad & v_y=\dot y,\\
v_\theta=\dot\theta,\qquad & v_\varphi=\dot\varphi,\\
\end{array}\right.
\end{equation*}
on $[t_0,t_1]-\left\{\tilde t\right\}$. 

\textbf{Conditions for the impact:} The impact condition at $t=\tilde t$ given in \eqref{eq:impactL} reads
\begin{equation*}
\left\{\begin{array}{l}
p_x^+-p_x^-=\lambda^1,\\
p_y^+-p_y^-=\lambda^0+\lambda^2,\\
p_\theta^+-p_\theta^-=-\lambda^1\,R\cos\varphi-\lambda^2\,R\sin\varphi,\\
p_\varphi^+-p_\varphi^-=\lambda^0\,R\cos\varphi,
\end{array}\right.
\end{equation*}
where we denote $p_x^+=p_x\left(\tilde t^+\right)$, etc., and $\lambda^0,\lambda^1,\lambda^2\in\mathbb R$ are the Lagrange multipliers. Similarly, the condition \eqref{eq:impactDelta} reads
\begin{equation*}
\left\{\begin{array}{l}
v_x^+=\lambda^3\,R\cos\varphi,\quad v_\theta^+=\lambda^3,\quad \\
v_y^+=\lambda^3\,R\sin\varphi,\quad v_\varphi^+=\lambda^4
\end{array}\right.
\end{equation*}
where $v_x^+=v_x\left(\tilde t^+\right)$, etc., and $\lambda^3,\lambda^4\in\mathbb R$ are the Lagrange multipliers.

For instance, when the disk hits the wall orthogonally, i.e., when $\varphi\left(\tilde t\right)=\pi/2$, the only admissible solution of the impact equations is
\begin{equation*}
\begin{array}{ll}
p_x^+=p_x^-=0,\quad  p_y^+=-p_y^-,\quad p_\theta^+=-p_\theta^-,\quad p_\varphi^+=p_\varphi^-.
\end{array}
\end{equation*}
\bibliography{ifacconf}



\end{document}